# Transient hydrophobic exposure in the molecular dynamics of Aβ peptide at low water concentration

Rukmankesh Mehra and Kasper P. Kepp*

*Technical University of Denmark, DTU Chemistry, Building 206, 2800 Kgs. Lyngby, Denmark.*

* Correspondence to Kasper P. Kepp, e-mail: kpj@kemi.dtu.dk

**ORCID of Authors**

0000-0001-6010-1514, 0000-0002-6754-7348

## Abstract

Aβ is a disordered peptide central to Alzheimer's Disease. Aggregation of Aβ has been widely explored, but its molecular crowding less so. The synaptic cleft where Aβ locates only holds 60-70 water molecules along its width. We subjected $A\beta_{40}$ to 100 different simulations with variable water cell size. We show that even for this disordered aggregation-prone peptide, many properties are not cell-size dependent, i.e. a small cell is easily justified. The radius of gyration, intra-peptide, and peptide-water hydrogen bonds are well-sampled by short (50 ns) time scales at any cell size. Aβ is mainly disordered with 0-30% α-helix but undergoes consistent α-β transitions up to 14% strand in 5-10% of the simulations regardless of cell size. The similar prevalence in long and short simulations indicate small diffusion barriers for structural transitions in contrast to folded globular proteins, which we suggest is a defining hallmark of intrinsically disordered proteins. Importantly, the hydrophobic surface increases significantly in small cells (confidence level 95%, two-tailed t-test), as does the variation in exposure and backbone conformations (>40% and >27 % increased standard deviations). Whereas hydrophilic exposure dominates hydrophobic exposure in large cells, this tendency breaks down at low water concentration. We interpret these findings as a concentration-dependent hydrophobic effect, with the small water layer unable to keep the protein unexposed, an effect mainly caused by the layered water-water interactions, not by the peptide dynamics. The exposure correlates with radius of gyration ($R^2 \sim 0.35-0.50$) and could be important in crowded environments, e.g. the synaptic cleft.

# Introduction

This paper used molecular dynamics (MD) simulations to explore the conformational ensembles of the intrinsically disordered peptide Aβ as a function of variable cell size and peptide concentration, which is both of technical and chemical interest. Aβ plays a central role in Alzheimer's disease (AD) as the constituent of the hallmark senile plaques in patient brains and as the toxic oligomer-forming species whose formation is affected by mutations in genes APP and PSEN1 and PSEN2 that cause severe early onset familial AD.[1]–[4] Understanding the dynamics and structural transitions of this peptide under various physiologically relevant conditions is of major interest, with a particular aim of targeting the pathogenic conformations by molecular intervention such as antibodies or anti-aggregation agents.[5]–[10]

Because of its structural disorder in solution, the peptide's toxicity and other modes of action are very conformation-dependent and the relevant conformations are very hard to relate to biological activity.[11]–[13] Classical MD simulations are massively used to study the structure and behavior of macromolecules,[14]–[18] and has served a key role in relating structural states of Aβ to its potential biological activity.[19]–[28] Despite these efforts, the bioactive conformations of Aβ in human brains remain unknown, although they probably involve hydrophobic exposed parts that correlate with toxicity[12], [25], [29] and may enhance oligomerization[10], [11], [30], [31] or interact maliciously with cell membranes.[32]–[35] Aβis known to concentrate at the synaptic cleft which represents one of the most important places in living systems as the center of neuronal transmission and also represents extremely crowded molecular environments, with a typical width of the cleft being 20 nm.[36], [37] From this we calculate that only 60-70 aligned water molecules separate the two synaptic terminals; in this massively crowded region Aβ is likely to display its main normal and pathogenic functions.[38]–[41]

On a technical note, the enormous phase space of atomic coordinates and velocities makes any simulation unlikely to be ergodic, i.e. to cover the interesting phase space, and many infrequent events (defined here as a transition between two conformations in phase space) may never be observed on the simulation time scale[42]. The "sampling problem" produces a concern whether most MD simulations are actually adequately averaged and perhaps a lack of trust in MD simulations[43]. Another concern of both fundamental and practical interest is the composition of the protein-water system itself. Even if the chemical model and force field reflect adequately the chemical state of the protein, the solvent-protein system is finite with a periodic simulation box used by most current MD simulations. The small finite size may introduce artifacts both due to periodicity and due to the erroneously high solute concentrations, with cell-size dependencies previously found for diffusion coefficients[44], [45] and thermal conductivities[46]. We are not aware of any previous investigation of the cell-size effect on the conformation of an intrinsically disordered protein.

Many studies have investigated the effect of Ewald summation and cut-offs of the electrostatic and the Lennard-Jones interactions[47]–[53]. Cutoff effects are, not surprisingly, particularly important for electrostatic interactions between atoms with large point charges, including charged residues of proteins[47], [52], [54], [55]. Artificial periodicity becomes particularly problematic as the Coulomb interaction increases, i.e. for highly charged, concentrated solutes in low-dielectric solvents[52]. The water interactions are impaired in small cells,[56] but many properties are independent of cell size,[57] whereas others are very cell-size dependent.[44] Given these variable findings, exploring the impact of variations in simulation cell sizes on disordered proteins should be of interest both to understand the fundamentals of the water-protein system and to potentially reduce computer time if the properties of interest are not size-dependent, considering that MD concentrations are much lower than the crystal structures but much higher than the experimental characterization assays, both routinely used in comparison.

To understand the impact of cell size on MD simulations, we study a small, charged protein that has a particularly context- and concentration-dependent ensemble. $A\beta_{40}$ is an intrinsically disordered peptide ideal for this purpose. It carries a charge of -3 at physiological pH and is extremely aggregation prone, and its secondary structure depends on the surroundings[58]–[60], plausibly enabling a conformation change from compact globular in solution to extended helical in membranes[11], [13], [61], [62]. The structural ensemble of $A\beta_{40}$ is important because its aggregation and membrane interactions are main culprits of disease[9], [10], [63]–[65], and exposed hydrophobic parts and coil character have been found to correlate with the toxicity of Aβ variants that cause disease[24], [25], [29]. Thus, not only for technical but also biological reasons the structure and dynamics of Aβ at high concentration is of considerable interest. Molecular crowding effects on protein conformation can arise for two reasons: Reduced water potential (concentration), or direct solute-solute interactions. We argue that it is important to separate the first generic effect from the latter effect and thus study here the "generic" crowding effect without any modulating influences of other solutes (dimers, trimers, other macromolecules) in the simulation cell, which would not be very generic.

We performed MD simulations of A$\beta_{40}$ using simulation cells of five different sizes (3 Å, 5 Å, 10 Å, 15 Å, and 20 Å). For each cell size, we performed 20 randomly seeded simulations for 100 nanoseconds, giving 2000 ns for each cell size. We also performed three additional longer MD simulations of 1000 ns each, to test the impact of averaging data for long vs. short MD simulations. Furthermore, we studied the peptide for additional 10 x 100 ns at 320 K to understand the sensitivity of our findings to reasonable variations in temperature. To understand the impact of cell size and increased crowding on the structure and dynamics of A$\beta_{40}$, we analyzed the secondary structure, radius of gyration, hydrophilic and hydrophobic solvent accessible surface area (SASA), inter- and intra-molecular hydrogen bonding.

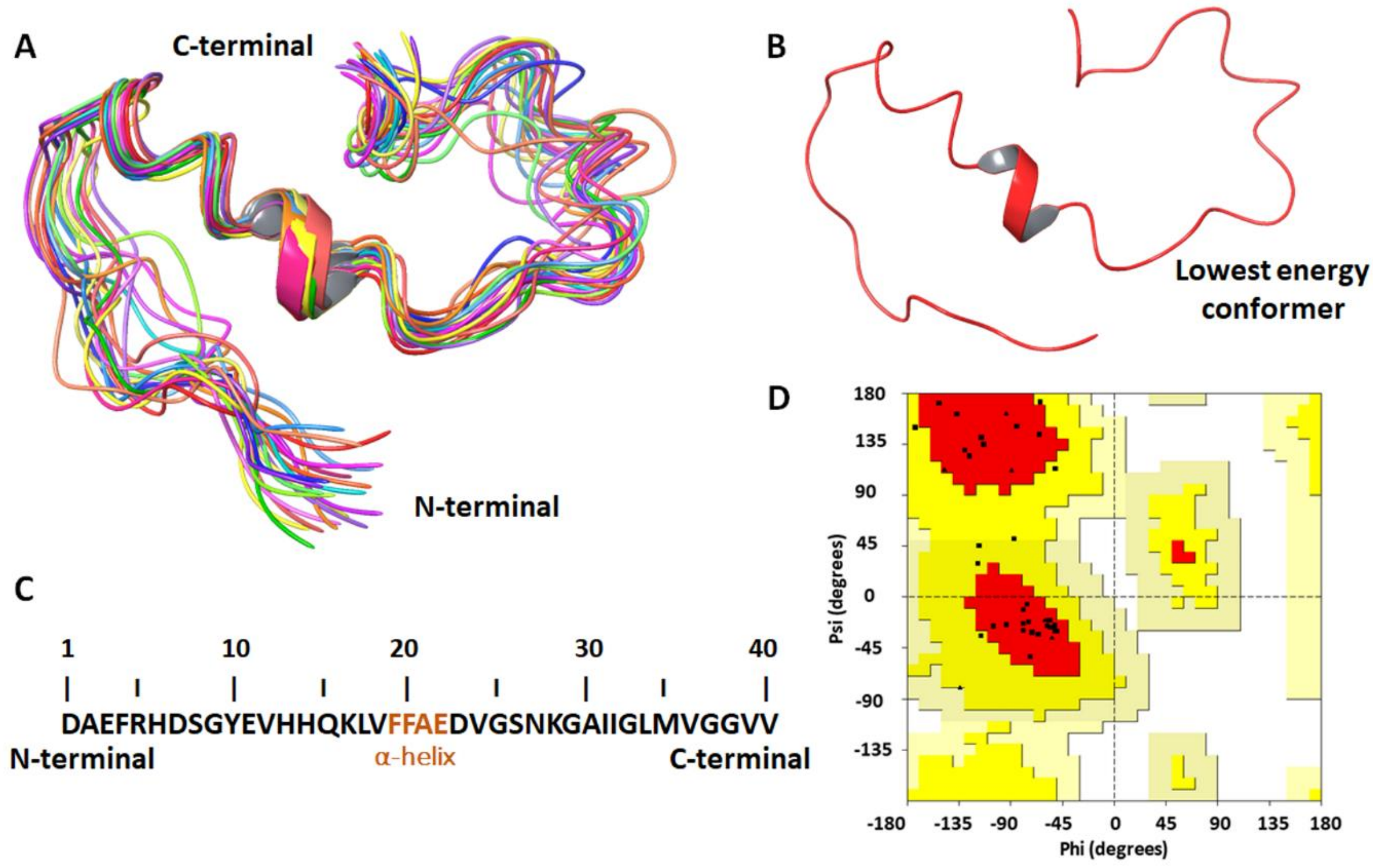

**Figure 1. A)** Structurally aligned twenty conformations of A$\beta_{40}$ from the PDB structure 2LFM. **B)** Lowest energy conformation (top) of A$\beta_{40}$ used in the present study. **C)** Amino acid sequence of A$\beta_{40}$ showing the residues forming α-helix. **D)** Ramachandran plot of the lowest energy conformation, with 25 residues in the core region (red) and seven residues in the additional allowed region (yellow). No residue is present in the disallowed region (white).

## Methods

### Starting A$\beta_{40}$ model

We studied a single solute rather than a dimer or oligomer of Aβ, because the multiple solutes cause aggregation and solute interactions that do not reflect the generic crowding effect of interest here. as the water potential increases the concentration decreases. An important factor is the interaction between periodic images between solutes in finite sized cells. These image interactions may cause artifacts, which are difficult to estimate but can also affect outcome.

The experimental NMR structure of A$\beta_{40}$ in aqueous solution (PDB code 2LFM; **Figure 1**)[58] was used as it represents the peptide in 100% water (other NMR-derived structures have co-solvents or co-solutes). This entry contains 20 conformations ranked based on lowest energies, with structural RMSD 1.7-3.7 Å (**Figure 1A**). The lowest energy conformation was used in the present study (**Figure 1B**). The peptide was prepared using Protein Preparation Wizard with default settings[66] at pH 7. Hydrogen atoms were added, hydrogen bonds were optimized and a local protein optimization was performed to remove steric clashes. The prepared protein contains -3 charge at physiological pH (please see the amino acids in the sequence, **Figure 1C**), with the N- and C-terminals charged as ammonium and carboxylate groups, respectively. The start structure is in good agreement with available experimental NMR and CD data and has a backbone conformation without disallowed regions (**Figure 1D**).

### System preparation

In order to understand the effect of the cell size in MD simulations, five different orthorhombic cells, ranging from very small to large, were studied viz. 3, 5, 10, 15 and 20 Å. These cell sizes are defined by using a buffer distance from the cell edge to the protein. Commonly used box sizes have a 10 Å buffer distance. Previously, we observed that a combination of Charmm22* force field[67] and TIP3P water model[68] produces conformational ensembles in excellent agreement with experimental NMR and CD data, which is important because of the major effect of force field on the structure of Aβ[69]. The more polar TIP3P water model tends to favor more compact folded structures, consistent with a larger hydrophobic effect of the stronger solute-solute TIP3P interactions vs. other water models[69]. Therefore, we prepared the systems using these force fields. Each system was neutralized by adding 3 $Na^+$ ions and additional 0.15 M NaCl was added in each. The force fields were applied using viparr.py and

build_constraints.py Python scripts of the Viparr tool and the systems were prepared using System Builder tool of Desmond[66], [70].

The five systems were prepared with their specific concentrations of $A\beta_{40}$ as listed in **Supplementary Table S1**. There is approximately 10 times more water molecules in the largest cell (20 Å, 10607 water molecules) than in the smallest cell (3 Å, 1097 water molecules). The concentrations range from about 0.005 M to 0.04 M, all of which are very high concentrations, with the latter corresponding to the highest realize concentration that we can simulate with the peptide still being in a solution. Please note that most MD simulations in the literature are in the middle range of this interval[69], whereas experimental assay concentrations are much more dilute, in the micromolar range[38].

**Energy minimizations and MD simulations**

Each system was energy-minimized using a combination of steepest descent and Broyden−Fletcher−Goldfarb−Shanno methods. MD simulation was then performed in NPT ensemble at 300 K and 1.0013 bar for 100 ns using a multistep protocol of Desmond[66], [70]. The integration time-step of 2 femtoseconds was used. The energies were noted with an interval of 1.2 picoseconds (ps) and trajectories were recorded at each 100 ps interval that generated 1001 snapshots in each run. The temperature and pressure were kept stable using the Nose-Hoover chain thermostat[71] and the Martyna-Tobias-Klein barostat[72]. The long-range electrostatic interactions were calculated using Ewald mesh summation technique with a cut-off of 9.0 Å. Each simulation performed in this study was initiated using a different, randomly seeded start velocity, producing in total 10 microseconds of concentration-dependent all-atom simulation of the peptide, divided into 100 seeded 100-ns simulations.

Additional ten randomly seeded simulations with 5 Å cell size were performed for 100 ns at 320 K with all other parameters similar to the other simulations, in order to analyze the effect of varying temperature on both sides of the physiological regime (most assays are performed at room temperature, 300 K, but the peptide is active at body temperature, 310 K). We hypothesized that thermal disorder could affect the peptide dynamics in particular in the high-concentration limit where the water-water interactions are weaker.

Finally, to understand the impact of sampling using many short (100 ns) vs. a few longer (1000 ns) MD simulations, we also performed three randomly seeded simulations of the system with 10 Å cell size, each lasting for 1000 ns at 300 K under same conditions as above except for the trajectory recording time. The trajectories were recorded at an interval of 1000 ps, producing a substantially longer averaging history relevant to the identification of slow processes with considerable diffusion barriers. Specifically, we wanted to test if there were any differences in the buildup of strand character at long simulation times to a level that could not be reached on the 100-ns time scale, as a very slow time scale of strand formation could prevent its observation in the shorter simulations despite their different seeds. Each 100 ns trajectory was analyzed for the equilibrated last 50 ns time, and the 1000 ns trajectories were analyzed for the last 950 ns and 500 ns time, to avoid artifacts of the first equilibration phase where structures diverge from the minimized initial start structure. To test for long diffusion-limited processes, we averaged the long simulations both over the last 950 ns and the last 500 ns and analyzed the differences compared to the short simulations.

The secondary structures, RMSD and RMSF plots were generated using the Simulation Interaction Diagram tool of the Desmond, and the hydrogen bonds and radius of gyration were studied using Simulation Event Analysis tool. For analyzing the hydrophobic and the hydrophilic SASA, the Python script trj_sasa.py was used with a probe radius of 1.4 Å, and the representative structures of the trajectories were identified by hierarchical clustering using the trj_cluster.py script of Schrodinger with default parameters[66].

# Results and discussion

## Sampling quality of MD simulations

To estimate whether MD simulations adequately sample the property of interest, one should compare many simulations with different start velocities, run these simulations for a considerable time, and then consider them together as independent observations[73]. Only the later horizontal parts of the trajectories are included in the statistics, as the initial trajectories reflect divergence from the starting structure. The RMSD and RMSF plots of all trajectories are shown in **Supplementary Tables S2-S13**. Unfortunately, in contrast to many systems and experiments where precision is usually quite high and triplicates are enough to reach decent results, Aβ is fundamentally disordered and has a large variance in its conformational ensemble, which produces stochastic (chaotic) results even in experimental assays.[38]

Similarly, very large standard deviations can be expected in MD for observables with large amplitude of motion even if they are well-sampled; this is true for conformational properties of Aβ as these are naturally fluctuating on fast timescales[69]. Some properties with slow time scales (> 1 microsecond) may be completely outside the simulation time scale, most notable changes in secondary structure of stable folded proteins, which have large diffusion barriers to change structure[74], [75]. Not surprisingly, in disordered proteins, the diffusion barriers to changes in secondary structure are much smaller. We show below that longer (1000 ns) and shorter simulations (100 ns) show the same prevalence of secondary structure transitions, and thus the characteristic time for these events is < 100 ns for this disordered peptide, much shorter than for compact folded proteins with large barriers separating the secondary structure types. Accordingly, as analyzed below, all the properties of interest in this study are well sampled on the applied simulation time scales.

## Disordered Aβ with ~0-27% (average 8-12%) helix and common α-β transitions

The analysis of the secondary structures in all the simulations at 300 K (with different cell sizes: 3, 5, 10, 15 and 20 Å) shows that α-helix was persistent except for a few simulations where β-strand was prominent (**Figure 2A-2D** and **Supplementary Tables S14-S20**). In all cases studied, the peptide remains predominantly disordered, with most of the residues being coil, turn or bend, in accordance with the experimental data.[58], [69] Complete loss (< 2%) of secondary structure, with neither helix or strand present, occurred regularly (13 times out of 100 simulations) and was independent of cell size.

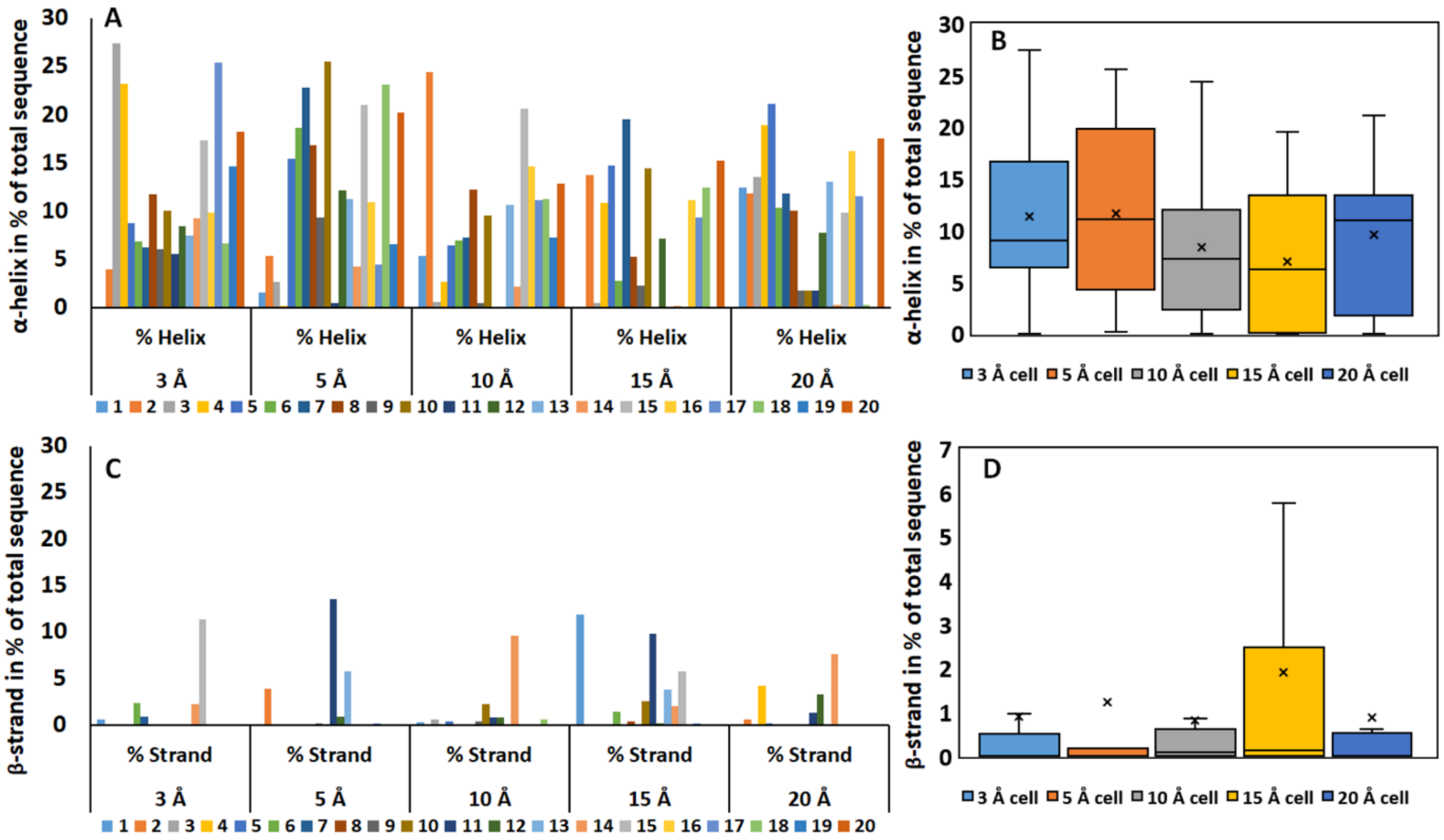

**Figure 2. A)** Percentage of residues with α-helix structure for 100 different 100-ns simulations of A$\beta_{40}$, with 20 simulations at five different cell sizes (3, 5, 10, 15, and 20 Å) (averaged over the last 50 ns of each simulation). **B)** Box-and-whisker plot of the helix content shown in A. **C)** Plot (as in A) of the percentage of β-strand. **D)** Box-and-whisker plot of strand content.

In most simulations, some helix was present, averaging to ~8-12% for the five cell sizes, with standard deviations of 6-8% helix character (**Supplementary Table S20**). The average strand character was 1-2%. Typically, β-strand formation to 5-15% was observed in 1-2 of each of the five sets of 20 seeded 100-ns simulations, making α-to-β transition a relatively common feature (5-10%) of the peptide at all concentrations. None of these tendencies were significantly dependent on the peptide concentration, and the similar tendency of α-to-β transition in all five groups suggest that the initial process of this transition is well-sampled, although the diffusion barrier and thus full strand content may not be so (as shown later from longer 1000-ns simulations, it is). In the 15-Å cell, two larger (10-12%) α-to-β transitions and two smaller (4-6%) produced a larger standard deviation (SD) in the strand content for this peptide but these variations are still less than 6% and should be seen in the context of the very small average strand content. In conclusion, the secondary structure and initial events of the α-to-β transitions are well-sampled and independent on simulation cell size.

### Hydrogen bonds and backbone conformations of concentrated Aβ in water

In order to analyze the hydrophobic packing of $A\beta_{40}$, intra-peptide and peptide-water hydrogen bond networks were analyzed (**Figure 3A,** and **Supplementary Table S20**). With an increase in the hydrophobic packing, the peptide-water hydrogen bond networks will decrease and intra-peptide hydrogen bonding will increase. For the 3-Å cell, the number of intra-peptide and peptide-$H_2O$ hydrogen bonds lies in the intervals 16.2−25.3 and 100.6−118.7, respectively, with averages of 20.0 (SD = 2.7) and 111.6 (SD = 5.0), respectively. In the 5-Å cell, the number of intra-peptide and peptide-$H_2O$ hydrogen bonds is 15.9−24.6 and 102.8−121.2 for the simulations, and their averages over all simulations are 20.7 (SD = 2.9) and 111.1(SD = 4.9), respectively. Similar patterns prevail for the 10-Å, 15-Å, and 20-Å cells. Correspondingly, the peptide-water hydrogen bonds lie between 105.1−122.0, 101.3−125.9 and 104.7−123.0, with averages of 114.8 (SD = 4.5), 115.3 (SD = 6.9) and 113.2 (SD = 5.5) for the 10-Å, 15-Å, and 20-Å cells, respectively. We thus conclude that the number of hydrogen bonds is unaffected by the molecular crowding.

To understand the impact of the reduction in water potential on the backbone conformations of the peptide, the Ramachandran plots of the representative structures of each simulation from cluster analysis were also analyzed (**Figure 3B** and **Supplementary Table S20**; excluding the six glycine residues and two terminal residues, as they can occur in any region of the plot). For comparison to experimental data, the 20 NMR conformers from the 2LFM entry[58] were also analyzed. These conformers have 71.9−90.6% of the residues (23-29 residues) in the core region and 9.4−28.1% (3−9 residues) of the 32 analyzed residues in the additional allowed region.

In most of the representative structures for all cell sizes, no residues had disallowed backbone conformations, except for few simulated structures, where one or two residues (two residues only in two cases) were present in the disallowed region. Notably, in the smallest cell (3 Å), the variations in the core and additional allowed regions were consistently larger, a feature that relates to the high variation in the hydrophobic SASA. The 15-Å cell was an exception, caused by the high frequency of α- to-β transitions. However, the Ramachandran values of all cells are in excellent agreement with the 20 NMR conformers of 2LFM. The average core regions for 3, 5, 10, 15 and 20 Å cells were 83.5% (SD = 10.1), 86.2% (SD = 5.8), 80.9% (SD = 7.0), 81.9% (SD = 7.4) and 84% (SD = 6.5), respectively.

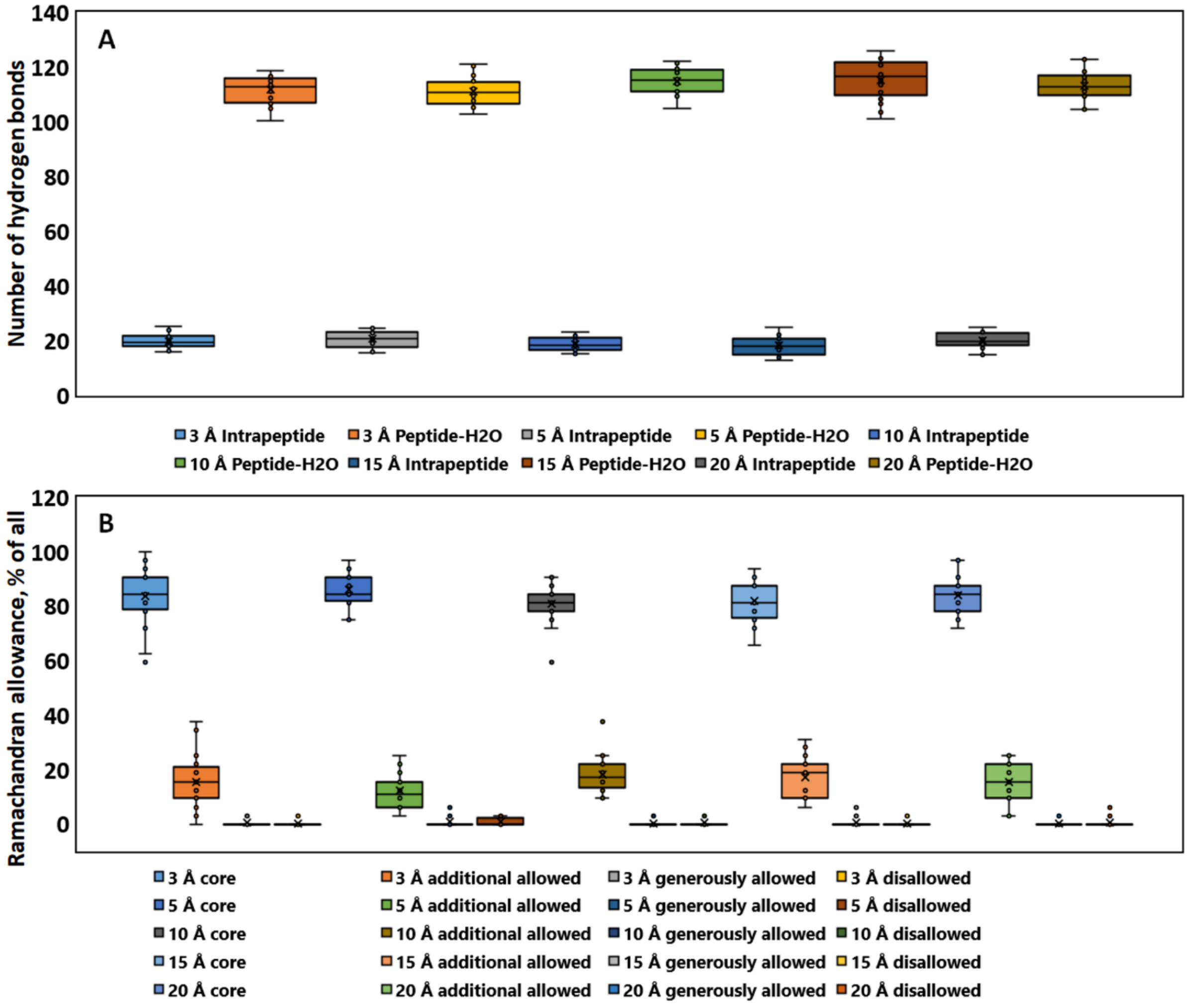

**Figure 3. A)** Box-and-whisker plot of the number of peptide-water hydrogen bonds and peptide-peptide hydrogen bonds for the last 50 ns of 100 different 100-ns MD simulations. The top values represent peptide-water hydrogen bonds and the bottom values represent peptide-peptide hydrogen bonds. **B)** Box-and-whisker plot of backbone conformations assigned to core, allowed, generously allowed and disallowed regions of the Ramachandran plot, averaged over the last 50 ns of all 100 100-ns simulations.

## Solvent accessible surface area

In order to understand the early conformational changes of A$\beta_{40}$ occurring at low water potential, the hydrophobic and hydrophilic solvent accessible surface area (SASA) of the peptide were analyzed for all 20 simulations of each of the five cell sizes (**Figure 4A** and **Supplementary Table S20**). It is interesting to note that Aβ has very similar total hydrophobic and hydrophilic exposure, both in the range 1500−2000 Å$^2$, indicating its very high amphilicity, a feature that rationalizes its well-established interaction with membranes[59], [61], [76].

We observe very interesting trends in the ensemble-averaged surface behaviors across the 100 simulations (**Figure 4A**). First, the hydrophobic SASA is significantly more variable in the two smallest cells where only a few layers of water are generally available. The average hydrophobic SASA in the small cells is also comparatively higher, 1760 $Å^2$ (SD = 271 $Å^2$) and 1718 $Å^2$ (SD = 272 $Å^2$) for the 3-Å and 5-Å cells, compared to the values 1598 $Å^2$ (SD = 186 $Å^2$), 1582 $Å^2$ (SD = 194 $Å^2$) and 1628 $Å^2$ (SD = 121 $Å^2$) of the 10-Å, 15-Å, and 20-Å cells (**Supplementary Table S20**). In all the larger cells, the hydrophilic surface area consistently separates from the hydrophobic surface area, whereas in the 3-Å and 5-Å cells, the hydrophobic exposure variability completely overlaps the range of hydrophilic exposure and thus routinely turn the balance in favor of hydrophobic exposure. We consider this observation very significant not only as a tangible physical-chemical consequence of the hydrophobic effect at low vs. high water potential, but also as a feature that controls the balance between the hydrophobic and hydrophilic tendencies of Aβ which are probably important to its bioactivity.

We conclude that Aβ has a smaller tendency to exhibit hydrophobic exposure at high water potential, and that the hydrophilic exposure outweighs the hydrophobic exposure, but that the amphiphilic balance is destroyed at low water potential. The differences are statistically significant, as shown from a two-tailed Student t-test for hypothesized same mean. Whereas the 3-Å and 5-Å cells have insignificantly different hydrophobic SASA (**Supplementary Table S21**), the mean hydrophobic exposure is significantly different at the 95% confidence level for the small vs. all the larger 10-, 15-, and 20-Å cells when these tests are performed independently (**Tables S22-S24**). In contrast, the hydrophilic surface is not significantly related to cell size (**Tables S25-S28**), and the average values are very similar with 1840 $Å^2$ (SD = 142 $Å^2$), 1846 $Å^2$ (SD = 117 $Å^2$), 1791 $Å^2$ (SD = 125 $Å^2$), 1816 $Å^2$ (SD = 163 $Å^2$), and 1779 $Å^2$ (SD = 119 $Å^2$) for the 3-Å, 5-Å, 10-Å, 15-Å, and 20-Å cells, respectively.

We hypothesized that this significant change in the balance between hydrophilic and hydrophobic exposure may be accompanied by a change in the size of the peptide. To test this, the radius of gyration ($R_g$) of $Aβ_{40}$ in all the simulations of different cell sizes was computed. $R_g$ is very similar ranging across all simulations from 9.4 to 13.8 Å with average $R_g$ ~11 Å and SD 0.6-1.0 Å (**Figure 4B** and **Supplementary Table S20**). The change in cell size does not have a statistically significant effect on $R_g$ of the peptide. However, the within-ensemble averages of $R_g$ and hydrophobic surface exposure correlate significantly, as the exposure is related to a partial increase in the size of the peptide, and for all five cell sizes the relation has $R^2$ values between 0.35-0.50 (**Figure 4C** and **Supplementary Figure S1**).

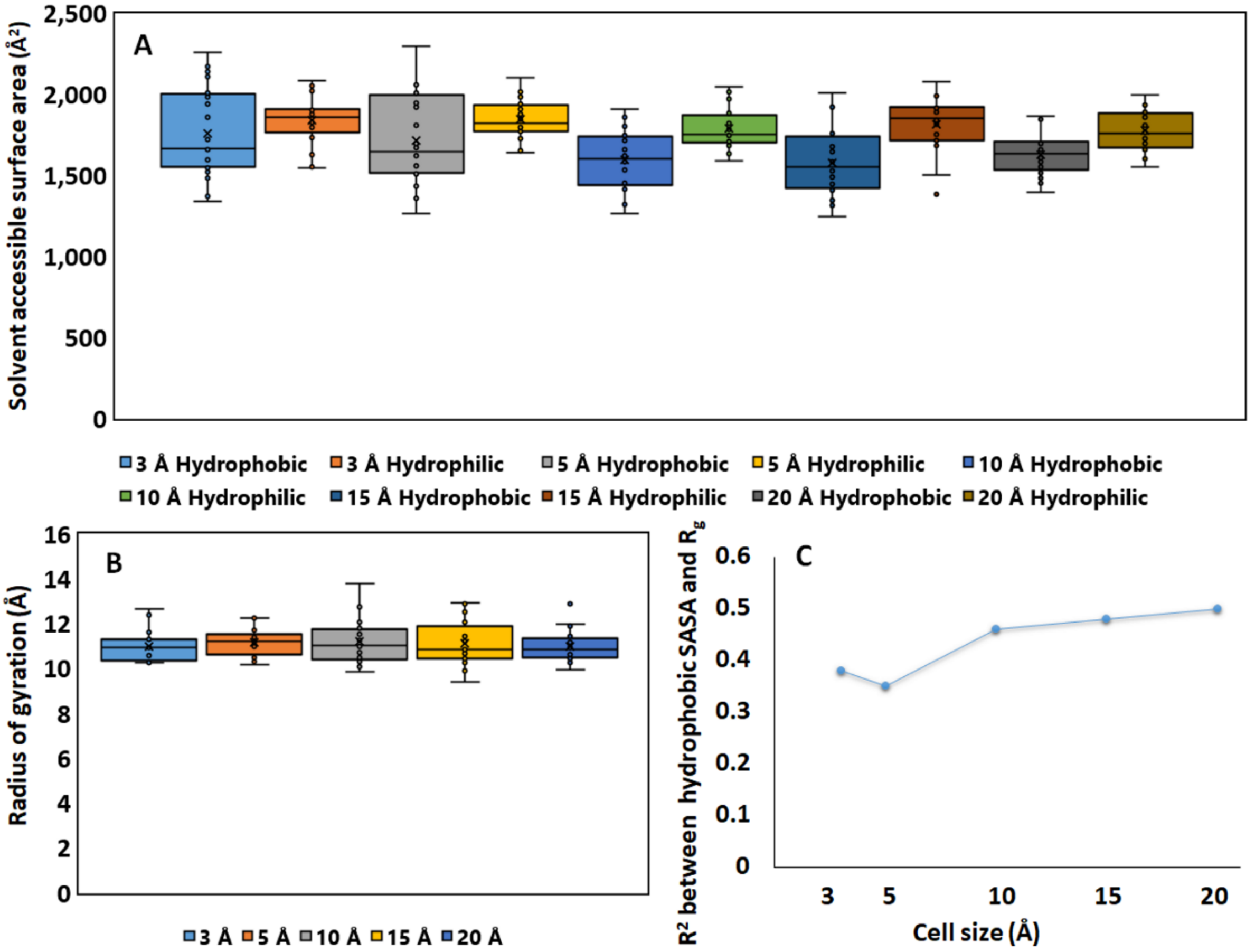

**Figure 4. A)** Solvent-accessible surface area of Aβ, divided into hydrophobic and hydrophilic parts, averaged over 20 MD simulations at five different simulation cell sizes (3, 5, 10, 15, and 20 Å). **B)** Box-and-whisker plot of the radius of gyration ($R_g$) of the simulations at five cell sizes. **C).** Calculated $R^2$ from linear regression of the relation between $R_g$ and hydrophobic solvent-accessible surface area for 20 simulations at each cell size; in all cell sizes, the two properties correlate significantly (95% confidence interval) with $R^2$ values of 0.35-0.50.

## Regional variations in secondary structure

The experimental NMR structure 2LFM contains 20 lowest energy conformers of Aβ$_{40}$.[58] All these conformers possess 10% α-helix (Phe-19 to Glu-22) (**Figure 1**) and no strand. In addition, they also comprise 15% $3_{10}$ helix (His-13 to Val-18), which are not very stable in water.[77] To avoid biases from the assignment, we focused our analysis on the most stable elements, i.e. all α-helices and β-strands. The percentage of α-helix in the experimental structure 2LFM is comparatively similar to the average α-helix for each cell size (**Figure 5A**), suggesting that our structural ensemble is in good accordance with the NMR data in water. Other NMR-derived structures of the Aβ monomer in PDB display more helix but they reflect the presence of co-solvents and micelles that increase the helix character[38]. Other data, notably coupling constants and chemical shifts for the peptide in pure water, also support that the peptide is mainly disordered with a small amount of helix and little strand.[69], [78] In contrast, the strand content rises quickly in the dimers and larger oligomers and is very large in the fibrils that form the senile plaques of AD[12], [79]

Apart from the α-helix in the experimental structure (Phe-19 to Glu-22), we also identified three more regions forming helices during the simulations (**Figure 5B, left** and **Supplementary Table S29**). We divided the residues into four helix-forming regions: Region 1 is from residue number ~1-10, region 2 is from ~11-20, region 3 is from ~21-26 and region 4 is from ~27-38 residues. Regions 3 and 4 were also prominent at all cell sizes, particularly region 4. Furthermore, as the cell size was increased up to 10 Å, the region 2 (experimental helix) occurred 7 times and region 4 occurred 11 times. Region 1 occurred very rarely at all cell sizes.

We also divided the residues into three strand regions: 1 (residues 1-10), 2 (residues 11-20), and 3 (residues 21-40) (**Figure 5B, right** and **Supplementary Table S29)**. All these regions occurred very rarely. In the 15-Å cell, β-strand occurred most frequently (11 times). Regions 1, 2 and 3 occurred three, four, and four times, respectively (**Figure 5B**). Thus, apart from the experimentally determined helix region[58], helix regions 3 and 4 have very high tendency to occur as α-helices; these are, interestingly, the regions that are experimentally observed in structures such as 1BA4 and other NMR-derived structures in the PDB where the water potential has been reduced[80]. Overall, we conclude that, as summarized by **Figure 5**, the helix and strand content of all simulations, even when divided into segments of the peptide sequence, are in good agreement with the experimental 2LFM conformations and are not significantly affected by cell size.

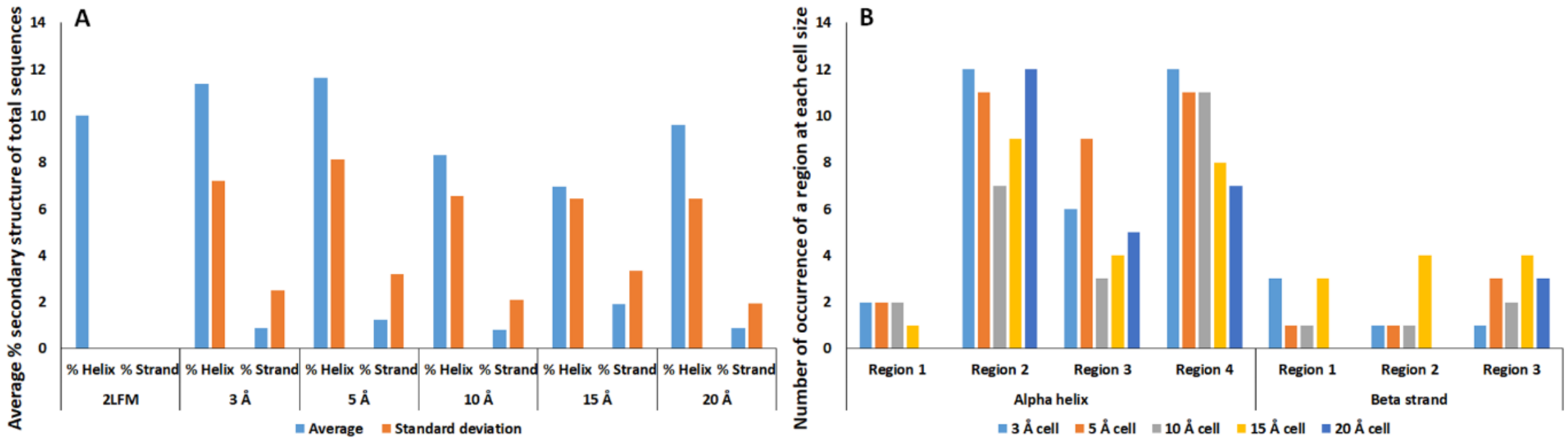

**Figure 5. A)** Average percent secondary structure found in each cell size compared with the average of 20 conformers of experimental NMR structure 2LFM. **B)** Analysis of the regions forming α-helices and β-strands. Number of times a region forms an α-helix or a β-strand in the 20 simulations of each cell size. The secondary structural regions that occurred ~ >20% of the analysis time (last 50 ns) of each run were analyzed.

## Results are not sensitive to variations around physiological temperature

The hydrophobic effect is related to temperature and manifests as an entropy-driven exposure of proteins and peptides as the water network is weakened. We thus hypothesized that the temperature could affect the observed exposure. To ensure that our findings are not sensitive to reasonable temperature changes near physiological temperature, we performed 10 x 100 ns of additional simulations for the second-smallest (5-Å) cell at 320 K, which is 10 K above physiological temperature, with the other simulations performed at 300 K, 10 K below body temperature, and typical for experimental assays of Aβ.

We compared this 320-K simulation with the 300-K simulation of the same 5-Å cell system (**Supplementary Table S20**). At 320 K, large variability is seen in all properties to the same extent as at 300 K. The increasing temperature in principle causes thermal disorder in the water-water interactions which could affect our conclusions, but the variability is similar at both temperatures and robustly captured by the averaging of the ensemble properties. $R_g$ was insignificantly different at 300 and 320 K: At 320 K, $R_g$ lies between 10.3-12.2 Å, with average 11.3 Å and SD 0.7 Å. The intra-peptide (average = 20.6, SD = 3.4) and peptide-water hydrogen bonds (average = 108.2, SD = 6.3) were insignificantly affected by the higher temperature. Most importantly, the increasing temperature had no effect on the hydrophobic SASA in the range around physiological temperature (**Supplementary Table S20**), and thus, our observations of a hydrophobic exposure at low water potential is robust against variations in temperature in the range relevant to assays and physiological conditions.

## Averaging data for long versus short simulations

It might be argued that some of the slowest modes in the ensemble, notably the α-to-β transitions, occur on longer timescales than those studied here, and that they require more time to travel up the diffusion barrier; this argument rests on the assumption that disordered peptides have slow diffusion barriers for these transitions as commonly seen for compact folded proteins[74]. However, intuitively, we expect disordered proteins to have much smaller barriers to these transitions and the question then emerges whether these transitions have a longer characteristic time scale. In order to answer this question, we compared averaging for long versus short MD simulations by performing additionally three 1000 ns simulations and comparing these to the twenty 100 ns simulations using the 10-Å cells. For the long simulations, the data were additionally averaged both over the last 950 ns and the last 500 ns to detect any timescale-dependence of the properties and their variations (**Figure 6**).

One long simulation produced more strand than the average of the 20 100-ns simulations. In this simulation, 5.2% and 9.9% strand occurred in the last 950 ns and 500 ns of the 1000 ns, indicating a buildup and persistence of strand character (**Figure 6A** and **Supplementary Figures S2** and **S3**). The percent α-helix in this long simulation was comparable to the average of the short simulations. 13.9% and 10% α-helix were found during the last 950 ns and 500 ns, compared to 8.3% average α-helix (SD = 6.5) in the short simulations. In contrast, the two other long simulations exhibited secondary structures very similar to the average of 20 short simulations. The percentage of α-helix and β-strand falls in the intervals 5.4−6.4% and 0.5−2.7% during the last 950 ns and 5.0−7.2% and 0.1−1.2% during the last 500 ns, respectively. The amount of strand buildup during the first long simulation is not higher than seen in the short simulations, and the variance in the properties is similar.

From this analysis, we conclude that α-to-β transitions are well-sampled to the same extent by the long and short simulations but still require a total simulation time of ~1000 ns to ensure its observation. These transitions occur 5-10% of the simulation time, build up to ~10% regardless of time scale, and thus have much small diffusion barriers than for larger folded proteins. This finding is important because it shows that the timescale and thus length of MD simulations required depends not only on the type of transition but also its free energy barrier, which can be very small for disordered proteins and peptides. In larger compact folded proteins, the secondary structure elements are longer and tend to pack in tertiary structure which stabilizes greatly the secondary structure and produces very large barriers for structure transitions, which is in direct contrast to our findings for Aβ.

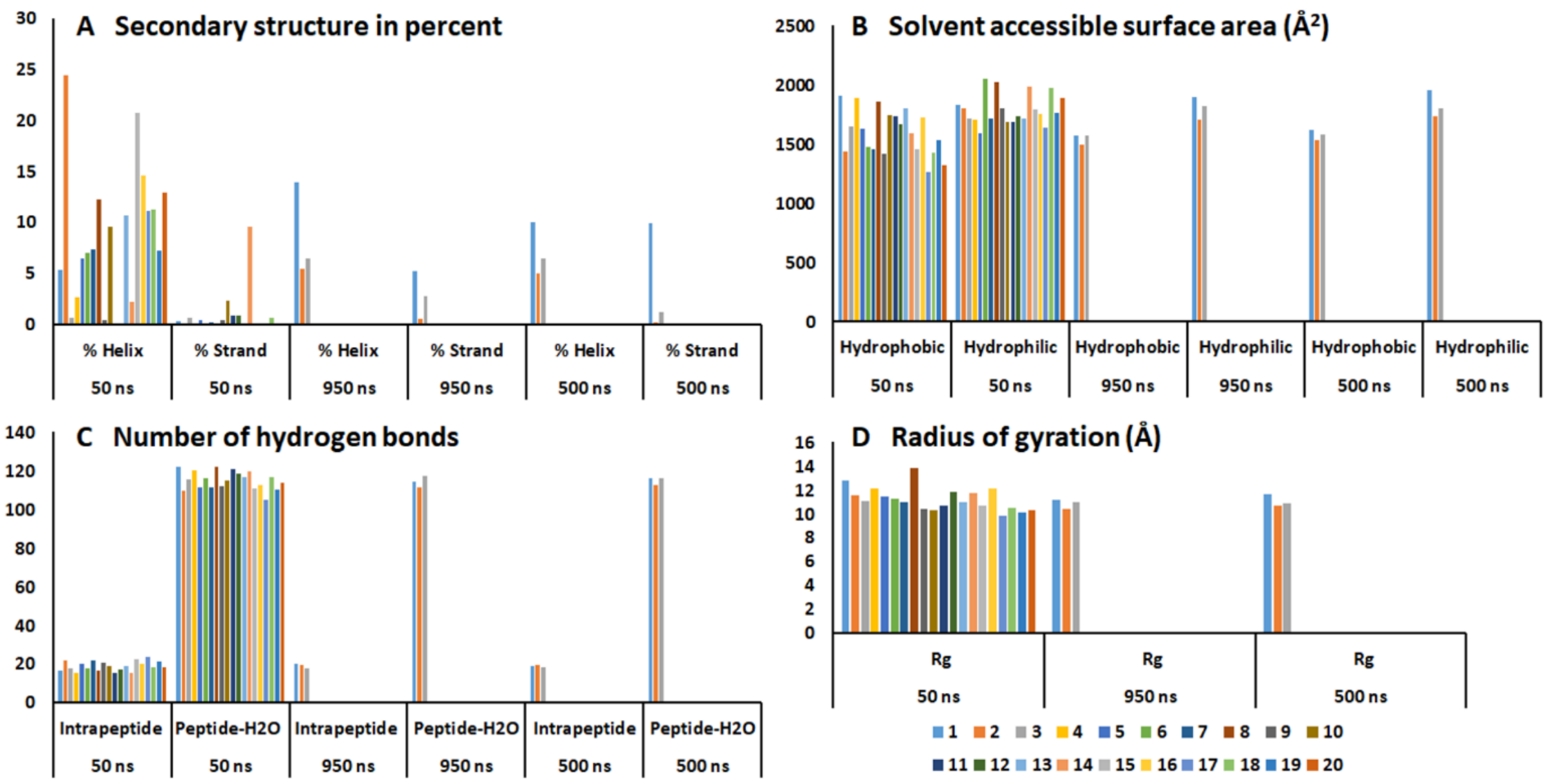

**Figure 6.** MD simulation results analyzed over 50 ns of the 100 ns runs (20 simulations) and over 950 ns and 500 ns of the 1000 ns runs (3 simulations). **A)** Secondary structure in percent. **B)** Hydrophobic and hydrophilic solvent accessible surface area (Å$^2$). **C)** Number of intra-peptide and peptide-water hydrogen bonds. **D)** Radius of gyration (Å).

The hydrophobic exposure is a much faster event but its prevalence could be biased by poorly sampled larger conformational changes. To support that our main conclusions on the hydrophobic exposure of Aβ are not dependent on the time scale of simulation, the hydrophobic and hydrophilic SASA averaged over the last 950 and 500 ns showed similar values to the average of the 20 100 ns simulations (analyzed for the last 50 ns, **Figure 6B**). For the 950 ns and 500 ns averaging, hydrophobic SASA lies between 1490-1571 Å$^2$ and 1531-1622 Å$^2$, and the hydrophilic SASA lies between 1705-1899 Å$^2$ and 1736-1956 Å$^2$, respectively.

We conclude that for intrinsically disordered peptide like Aβ (not for compact proteins undergoing secondary structure transitions in general) averaging over longer or many shorter MD simulations makes no significant difference because of the small transition barriers, whereas the total simulation time does matter and should reach at least 1000 ns for any system type (cell size, in our case).

## Conclusions

We show using the structure-balanced force field Charmm22*[67] that even for an extremely disordered and context-dependent peptide like Aβ, the cell size is unimportant for many properties typically studied by MD simulations. This includes the radius of gyration, hydrophilic SASA, intra-peptide and peptide-water hydrogen bonding, and secondary structure. All commonly performed MD simulations are extremely concentrated compared to experimental assay conditions, and in addition to sampling and force field quality, the chemical composition may thus be a major issue in MD simulations. We argue that it is not: Since Aβ is very disordered, we expect the same size-insensitivity to manifest for most other peptides. By comparing long and short simulation and different averaging, and using statistical tests for means, we show that our simulations sample all the properties well and thus recover the intrinsic variability of the peptide, which is very large, rather than variability due to statistical uncertainty. Our simulated structures are in excellent agreement with the NMR-derived conformations in pure water reflected by the structure 2LFM.[58]

However, one particular property, the hydrophobic SASA, was significantly affected by cell size: The hydrophobic exposure was significantly larger in small cells (3 and 5 Å). Similarly, higher dynamic variability was observed for the hydrophobic exposure and the backbone conformations as estimated from the Ramachandran regions. Although $R_g$ is not significantly affected by cell size, the exposure at low water potential correlates with $R_g$. We interpret the hydrophobic exposure at low water potential as a manifestation of the hydrophobic effect, with the main difference between large and small cells being the strength of the water-water interactions; in the small cells with only 1-2 layers of water in some areas around the peptide significantly impair the water-water interactions and consequently, the hydrophobic packing of the peptide. Hydrophobic exposure in genetic variants of Aβ are known to correlate with experimentally determined cell toxicities of the peptides[24], [25], [29].

The most important findings are that i) Aβ and probably other intrinsically disordered peptides have very small barriers for structural transitions such that they are sampled equally well many short vs. a few longer simulations; ii) most properties of interest are not cell-dependent and, considering the concentrated protein crystal structures and the dilute assays usually compared to MD data, it is justified to use a cell of 5 Å which will save considerable computer time; iii) However, for Aβ, the hydrophobic surface exposure is significantly larger in the very small cells (two-tailed t-test, 95% confidence). We interpret this as the hydrophobic effect disappearing in the limit of small water potential; thus, the hydrophobic effect in this system is driven by water-water interactions rather than the peptide dynamics. This type of exposure is probably biologically important because it correlates with cell toxicity, and thus, the identified exposure in parts of the peptide could be a relevant therapeutic target.

## Acknowledgements

The Danish Council for Independent Research | Natural Sciences (DFF), grant case 7016-00079B, is acknowledged for supporting this work.

## Supporting information available

The supporting information file contains all RMSD plots, RMSF plots and secondary structure plots for the simulations, including tables and figures analyzing the statistics of the ensembles. "**Supplementary Table S20**" is provided as a separate excel document containing data of secondary structure, radius of gyration, SASA, intra-peptide and peptide-water hydrogen bonds and Ramachandran plot analysis.

## Table of content graphic

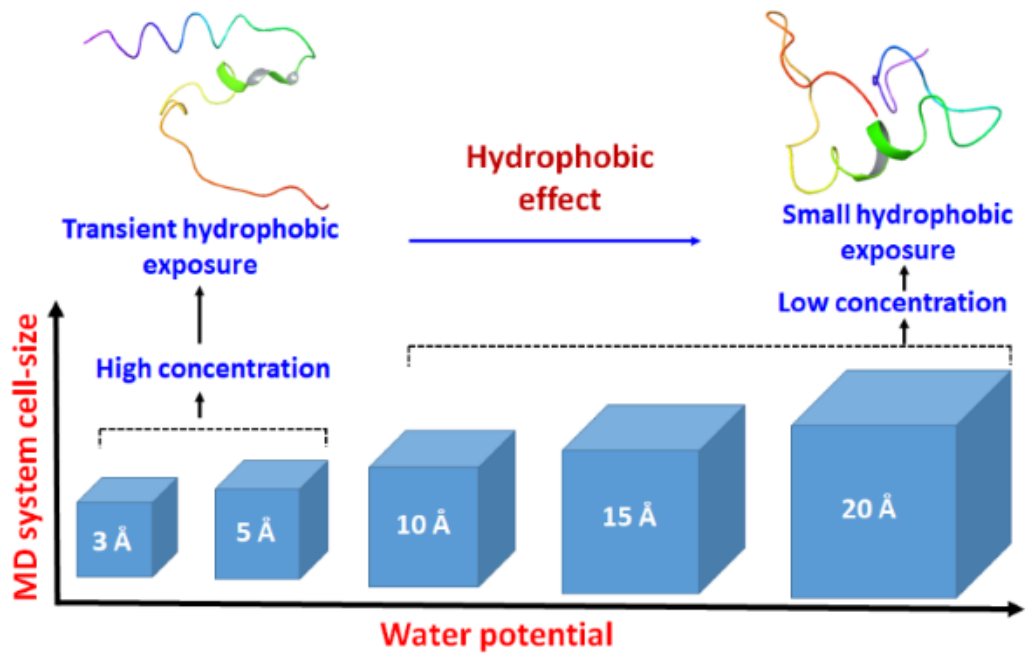